\documentclass[12pt]{article}
\usepackage{amsfonts,amssymb,epsfig,amsmath}
\usepackage{bbold}
\usepackage{slashed}
\usepackage{color}
\usepackage{hyperref}
\renewcommand{\text}[1]{#1}

\newcommand{\be}{\begin{equation}}
\newcommand{\ee}{\end{equation}}
\newcommand{\ben}{\begin{displaymath}}
\newcommand{\een}{\end{displaymath}}
\newcommand{\bea}{\begin{eqnarray}}
\newcommand{\eea}{\end{eqnarray}}
\newcommand{\bean}{\begin{eqnarray*}}
\newcommand{\eean}{\end{eqnarray*}}
\newcommand{\nn}{\nonumber \\}
\newcommand{\ba}{\begin{array}}
\newcommand{\ea}{\end{array}}
\newcommand{\bi}{\begin{itemize}}
\newcommand{\ei}{\end{itemize}}

\renewcommand{\theequation}{\arabic{section}.\arabic{equation}}
\def\theequation{\thesection.\arabic{equation}}

\def\l{\lambda}
\def\a{\alpha}

\def\b{\beta}
\def\g{\gamma}
\def\G{\Gamma}

\def\G{\Gamma}
\def\g{\gamma}
\def\e{\epsilon}
\def\s{\sigma}
\def\e{\epsilon}

\def\CE{\mathcal{E}}

\def\m{\mu}

\DeclareMathOperator{\vol}{vol}

\begin{document}

\makeatletter
\renewcommand{\theequation}{\thesection.\arabic{equation}}
\@addtoreset{equation}{section}
\makeatother

\begin{titlepage}

\vfill
\begin{flushright}
FPAUO-12/04\\
\end{flushright}

\vfill

\begin{center}
   \baselineskip=16pt
   {\Large \bf Self-duality of the D1-D5 near-horizon}
   \vskip 2cm
    Eoin \'O Colg\'ain
       \vskip .6cm
             \begin{small}
      		 \textit{Departamento de F\'isica, 
		 Universidad de Oviedo, \\
33007 Oviedo, ESPA\~NA}
             \end{small}\\*[.6cm]
\end{center}

\vfill \begin{center} \textbf{Abstract}\end{center} \begin{quote}
We explore fermionic T-duality and self-duality in the geometry $AdS_3 \times S^3 \times T^4$ in type IIB supergravity. We explicitly construct the Killing spinors and the fermionic T-duality isometries and show that the geometry is self-dual under a combination of two bosonic $AdS_3$ T-dualities,  four fermionic T-dualities and either two additional T-dualities along $T^4$ or two T-dualities along  $S^3$. In addition, we show that the presence of a $B$-field acts as an obstacle to self-duality, a property attributable to S-duality and fermionic T-duality not commuting. Finally, we argue that fermionic T-duality may be extended to $CY_2 = K3$, a setting where we cannot explicitly construct the Killing spinors. 

\end{quote} \vfill

\end{titlepage}

\section{Introduction}
A few years ago fermionic T-duality \cite{fermTdual,Beisert}, through its pivotal role in exhibiting the self-duality of $AdS_5 \times S^5$, offered a beautiful take on symmetries observed in the scattering amplitudes of $N=4 $ super Yang-Mills \cite{Alday:2007he,Drummond:2007aua,Brandhuber:2007yx} . In one dimension lower, a wealth of evidence \cite{Agarwal:2008pu,Bargheer:2010hn,Lee:2010du,Henn:2010ps,Huang:2010qy,Gang:2010gy,Bianchi:2011rn,Bianchi:2011dg,Chen:2011vv,Bianchi:2011fc,Wiegandt:2011uu,Bianchi:2011aa} for similar symmetries in the amplitudes of $N=6$ ABJM theory \cite{ABJM}  theory, suggests that $AdS_4 \times CP^3$ should also permit a self-dual mapping under a combination of bosonic and fermionic T-dualities. Despite some concensus on what form this combination should take \cite{Bargheer:2010hn}, neglecting the pp-wave limit  where self-duality is a more modest affair \cite{Bakhmatov:2011aa} (see also \cite{Bakhmatov:2009be}), a concrete geometric manifestation of these symmetries has been elusive \cite{Adam:2009kt,Grassi:2009yj,Adam:2010hh,Bakhmatov:2010fp,Dekel:2011qw}. This conundrum aside, given our preconceptions built on a staple of bosonic T-duality \cite{Buscher}, fermionic T-duality continues to challenge us and remains the subject of other recent studies \cite{Sfetsos:2010xa,Godazgar:2010ph,ChangYoung:2011rs,Grassi:2011zf, Nikolic:2012ih}. 

Recall that fermionic T-duality generates new ``solutions" by identifying commuting fermionic isometry directions built from Killing spinors \cite{fermTdual}. The requirement that the directions commute leads to the Majorana condition being compromised, and as such, solutions to complex supergravity ensue when one cranks the handle. However, in tandem with timelike T-duality, a process where the RR fluxes are complexified \cite{Hull:1998vg}, one can sometimes coerce the geometry back to its original form. Building on \cite{Bakhmatov:2009be,Bakhmatov:2010fp}, this work hopes to shed more light by constructing further examples based on a supergravity treatment using the Killing spinors. We confine our interest to the near-horizon of D1-D5 dual to $D=2$ SCFTs preserving $N=(4,4)$ supersymmetry. We will be interested primarily in the $T^4$ moduli space, but will also touch on $K3$.  

We start by showing that the geometry $AdS_3 \times S^3 \times T^4$ is self-dual under a combination of two bosonic T-dualities along $AdS_3$ and four fermionic T-dualities. While this may be expected from the $\sigma$-model analysis \cite{Adam:2009kt,Dekel:2011qw}, one slight twist in the expected story is that this leads to intersecting D3 branes i.e. the geometry is sourced by five-form fluxes. Recovering the original solution then boils down to two further T-dualities along $T^4$. Furthermore, we show that it is not possible to recover the geometry on the nose unless one resorts to additional T-duality along $T^4$. 

This work also serves to back up the instructive supergravity calculation presented in \cite{Bakhmatov:2010fp}. While one encounters a singularity in the dilaton shifts in the setting of $AdS_4 \times \mathbb{C} \textrm{P}^3$, here we see that one can explicitly perform $AdS_3$ and $S^3$ bosonic T-dualities by complexifying the $S^3$ as initially suggested in \cite{fermTdual}. We will see that the requirement that the Killing spinors are invariant under the spinorial Lie derivative \cite{hep-th/9902066} w.r.t. the Killing directions on a complexified $S^3$ singles out four Killing spinors by imposing a further projection condition. Somewhat interestingly, in contrast to the previous example, when one incorporates the additional $S^3$ bosonic T-dualities, one recovers the geometry without recourse to $T^4$ T-duality. This suggests that if $AdS_4 \times \mathbb{C} \textrm{P}^3$ is indeed self-dual that the methods employed in \cite{Bakhmatov:2010fp} may be put to use. The important point would appear to be  correctly identifying the appropriate complexification and the internal bosonic T-dualities so as to avoid singularities.  

The $AdS_3 \times S^3 \times T^4$ spacetime also allows us to touch upon S-duality. As there is a one parameter family of solutions with zero axion and dilaton that rotates the RR two-form potential $C^{(2)}$ and the NS $B$-field, one can ask if the geometry remains self-dual with a $B$-field. An initial hint that something may be amiss comes from studying the geometry supported solely by the $B$-field. In addition to finding a singularity in the bosonic $AdS_3$ transformations, one also sees that the fermionic T-dualities become trivial. Turning back on $C^{(2)}$, one discovers that fermionic T-duality leads to real and pure imaginary fluxes, where the latter are linked to the $B$-field and disappear once it is set to zero. Thus, the $B$-field is an obstacle to self-duality and suggests that fermionic T-duality is unlikely to have an anologue in an environment preserving no supersymmetry such as \cite{Lu:2011nga}. 

Finally, we close by reviewing our calculations and asking what they tell us about the D1-D5 near-horizon where $CY_2$ is $K3$ instead of $T^4$. By making use of the projection conditions, we show that the auxiliary matrix defining the fermionic T-duality does not depend on the $CY_2$. We then present an argument that the reshuffling of the fluxes is also independent of the internal $CY_2$ and therefore applies equally well to $K3$. 

Having spelled out the format of this work, we start by reviewing some essentials. 

\section{Background}
We begin this section by introducing fermionic T-duality in type II supergravity. We follow the conventions of \cite{Bakhmatov:2011aa,Bakhmatov:2009be} which we summarise in the appendix. In particular we will use a definite choice of Majorana-Weyl representation in which the $D=10$ gamma matrices are all real. In our conventions, given two Majorana-Weyl Killing spinors in type IIB, 
\be
\eta = \left( \begin{array}{c} \e \\ \ \hat{\e} \end{array} \right),  
\ee
the fermionic T-duality of Berkovits and Maldacena \cite{fermTdual} may be deduced from a single auxiliary matrix $C_{ij}$ satisfying
\be
\label{C}
\partial_{\mu} C_{ij} = 2 i \e_{i} \g_{\mu} \e_j, 
\ee
where $\e_i$ are Killing spinors subject to a spinor constraint\footnote{These relations are naturally enough independent of the choice of representation. Starting from the IIB formulation of Schwarz where complex spinors appear, it is possible to show that these conditions arise by following the approach of \cite{Godazgar:2010ph}.},
\be
\label{commutespin}
\e_i \g_{\mu} \e_j + \hat{\e}_i \g_{\mu} \hat{\e}_{j} = 0.
\ee
Here $i=1,\cdots,n$ runs over the number of commuting fermionic isometries with respect to which one is T-dualising. As was pointed out in \cite{Godazgar:2010ph}, this latter constraint comes as a result of integrability from (\ref{C}) and may be derived by acting again with the Killing spinor equation. In the complementary language of Berkovits and Maldacena this constraint corresponds to the requirement that fermionic isometry directions commute. 

In contrast to bosonic T-duality, the NS metric and $B$-field do not change when one performs a fermonic T-duality and the only change in the geometry is a shift in the dilaton 
\be
\label{dilshift}
\tilde{\phi} = \phi + \frac{1}{2} \textrm{Tr}(\log C), 
\ee
and a reshuffling of the fluxes 
\be
\label{fluxtrans}
\frac{i}{16} e^{\tilde{\phi}} \tilde{F}^{\a \b} = \frac{i}{16} e^{\phi} F^{\a \b} + C_{ij}^{-1} \e^{\a}_i \otimes \hat{\e}^{\b}_j,
\ee
where $F^{\a \b}$ is a type IIB bispinor: 
\be
\label{fluxbi}
F^{\a \b} = (\g^{\mu})^{\a \b} F_{\mu} + \frac{1}{3!} (\g^{\mu_1 \mu_2 \mu_3})^{\a \b} F_{\mu_1 \mu_2 \mu_3} + \frac{1}{2} \frac{1}{5!} (\g^{\mu_1 \mu_2 \mu_3 \mu_4 \mu_5})^{\a \b} F_{\mu_1 \mu_2 \mu_3 \mu_4 \mu_5}.
\ee
An explicit expression for the complementary bispinor in type IIA maybe found in \cite{Bakhmatov:2011aa}. It is these transformations that allow one to undo the effects of performing bosonic T-duality along $AdS_5$ and in the process show that $AdS_5 \times S^5$ is self-dual \cite{fermTdual}. Finally, observe also that chirality does not change, so the effect of fermionic T-duality is confined exclusively to either type IIB or type IIA. In some sense, this feature has overlap with non-Abelian T-duality which also permits one to preserve chirality provided the number of generators of the non-Abelian isometry group is even \cite{Sfetsos:2010uq, Lozano:2011kb}. 

Having glanced over the nuts and bolts of fermionic T-duality in type II, we introduce our subject matter. We will be interested in the near-horizon of D1-D5, namely $AdS_3 \times S^3 \times CY_2$. This geometry can be supported by either a two-form RR potential $C^{(2)}$, or a NS $B$-field, with $S$-duality connecting these two equivalent descriptions. Indeed, there is a one parameter family of solutions where both fields are turned on. Alternatively, this geometry may be supported by five-form fluxes corresponding to D3-branes wrapping a K\"ahler two-cycle in $CY_2$ (for example see \cite{Gauntlett:2007ph}). These equivalent descriptions are connected via either T-duality on $T^4$, or mirror symmetry (see for example \cite{Aspinwall:1996mn}) when $T^4$ is replaced by $K3$. 

Having introduced the various guises of the D1-D5 near-horizon, we turn our attention the simplest case of a geometry sourced by a RR three-from $F^{(3)}$, the preserved Killing spinors and fermionic T-duality. 

\subsection{Geometry with RR-flux} 
In type IIB supergravity, the $AdS_3 \times S^3 \times T^4$ solution may be written as
\bea
\label{soln}
ds^2 &=& ds^2(AdS_3) + ds^2(S^3) + ds^2(T^4), \nn
F_3 &=& 2 \left[ \vol(AdS_3) + \vol(S^3) \right],
\eea
where we have adopted the usual normalisations $R_{\mu \nu} = -2 g_{\mu \nu}$ for $AdS_3$ and $R_{\a \b} = 2 g_{\a \b}$ for $S^3$. Here, as is customary for fermionic T-duality with $AdS$ spacetimes, we write the metric on $AdS_3$ in terms of the Poincar\'e patch
\be
\label{poinmet}
ds^2(AdS_3) = \frac{-dt^2 + dx^2 + dr^2}{r^2},
\ee
and the metric on $S^3$ as
\be
\label{sphmet}
ds^2(S^3) = d \theta^2 + \sin \theta^2 (d \phi^2 + \sin^2 \phi d \psi^2).
\ee
With these metrics, a natural orthonormal frame may then be written
\bea
\label{vielbein}
&&e^{0} = \frac{dt}{r}, \quad e^{1} = \frac{dx}{r}, \quad e^{2} = \frac{dr}{r}, \nn
&&e^{3} = d \theta, \quad e^{4} = \sin \theta d \phi, \quad e^{5} = \sin \theta \sin \phi d \psi.
\eea

In performing fermionic T-duality, one needs to first identify the Killing spinors preserved by the supergravity solution. The Killing spinors for this geometry have already been worked out in \cite{Raeymaekers:2006np}. In the notation of \cite{hassan, Grana:2002tu}, the Killing spinor equations may be expressed in string frame as 
\bea
\label{dilatino}
\delta \lambda &=& \frac{1}{2}\G^M \partial_M \phi \eta- \frac{1}{4} \slashed{H} \sigma^3 \eta- \frac{1}{2} e^{\phi} \left[\slashed{F}_{1} (i \s^2) + \frac{1}{2} \slashed{F}'_{3} \s^1 \right]  \eta \\
\label{KSE} \delta \Psi_{M} &=& \nabla_{M} \eta -\frac{1}{8} \G^{NP} H_{MNP}  \s^3 \eta \nn &+& \frac{1}{8 } e^{\phi} \left[ \slashed{F}_1 \G_M (i \s^2) + \slashed{F}_3' \G_M \s^1 + \frac{1}{2} \slashed{F}_5' \G_M (i\s^2) \right] \eta,
\eea
where $F_{3}' = F_{3} - C_{0} H_3$ and $F_5' = F_5 - H_3 \wedge C_2$. 
As further explained in the appendix, we work with a real Majorana-Weyl representation of the $D=10$ gamma matrices, with $\G^0$ antisymmetric and all other gamma matrices symmetric. As we are working in the context of type IIB supergravity, the Killing spinor $\eta$ consists of two chiral spinors of the same chirality, $\G_{11} \e = \e$ and $\G_{11} \hat{\e} = \hat{\e}$,  where $\G_{11} \equiv \G^{0} \dots \G^{9}$. The Pauli matrices in the above Killing spinor equations act on $\e$ and $\hat{\e}$. 

While we will work with Killing spinors directly in $D=10$, it is also possible to decompose the Killing spinors in a $(6,4)$ split to distill off the $CY_2$. For the interested reader we enclose an appendix adopting this approach though caution that it will play no role in future discussions. 

Returning to the solution of interest (\ref{soln}) where only the three-form flux is non-zero, the vanishing of the dilatino variation (\ref{dilatino})  gives us the projection conditions
\be
\label{adssproj}
\G^{012345} \eta = -\eta, \quad \G^{6789} \eta = -\eta, 
\ee
where the second projection condition is not independent and follows from the first via chirality. As a result of imposing this condition, the supersymmetry is broken by one half. 

Along the $T^4$ direction, the Killing spinor is covariantly constant $\nabla_m \eta = 0$. The gravitino variation in the remaining directions may then be written in the form
\be
\left[ \nabla_{\mu} + \frac{1}{4} \left( \G^{012} + \G^{345} \right) \G_{\mu} \s_1 \right] \eta = 0,
\ee
where $\nabla_{\mu} \equiv \partial_{\mu} + \tfrac{1}{4} \omega_{\mu\rho\sigma} \G^{\rho \sigma}$. This equation may be solved for Poincar\'e Killing spinors to give 
\be
\label{killspin}
\eta = r^{-1/2} e^{-\tfrac{\theta}{2} \G^{4 5} \s^1} e^{\tfrac{\phi}{2} \G^{3 4}} e^{\tfrac{\psi}{2} \G^{4 5}}\eta_0,
\ee
where the constant spinor $\eta_0$ satisfies the projection condition for Poincar\'e Killing spinors
\be
\label{poinproj}
\G^{01} \s^1 \eta_0  = \eta_0.
\ee
Note we have neglected the superconformal Killing spinors and have just focused on the Poincar\'e Killing spinors, i.e. those depending solely on the radial direction of $AdS$. These are typically the Killing spinors that are most useful in exhibiting self-duality as in contrast to the superconformal Killing spinors (see \cite{Lu:1998nu}), their explicit form is simpler. 


As $(\e, \hat{\e})$ are Weyl, and $\G^{11} = \s^3 \otimes \mathbb{1}$ in our gamma matrix representation, they may simply be regarded as 16 component spinors comprising the upper 16 components of 32 component spinors. We may then rewrite (\ref{poinproj}) as (see appendix)
\be
\label{16dproj}
\g^1 \e = \hat{\e}.
\ee
This then determines $\hat{\e}$ in terms of $\e$ and all attention can shift to $\e$. Note that all the 16 dimensional gamma matrices are symmetric.

From (\ref{killspin}) we can then write
\be
\label{killspin2}
\e = r^{-1/2} e^{-\tfrac{\theta}{2} \g^{145} } e^{\tfrac{\phi}{2} \g^{34}} e^{\tfrac{\psi}{2} \g^{45}} \e_0,
\ee
where $\e_0$ denotes a constant spinor that we will expand in a basis of spinors below. In deriving this expression we have used the fact that $\hat{\e}_0 = \g^{1} \e_0$. Also it is worth reinforcing that this expression also determines $\hat{\e}$ given $\e$ and these can be combined into a pair of spinors $(\e, \hat{\e})$ corresponding to a fermionic isometry direction. 

Now that we have determined the form of the Killing spinor and extracted a pair $(\e, \hat{\e})$, in order to perform multiple fermionic T-dualities, we need to identify commuting fermionic isometry directions in tune with (\ref{commutespin}). 
Recalling again (\ref{16dproj}), we see that (\ref{commutespin}) holds if $\e_i \g^{1} \e_j = 0$ and $\e_i \e_j =0$ as $\g^{0}$ is simply the identity. This latter condition then tells us that the spinors we choose $\e_i$ should be complex. 

To find these commuting Killing spinors, we make use of a complex basis respecting the projector (\ref{adssproj}). Since $\e$ and $\hat{\e}$ are of definite chirality imposing $\G^{6789} \eta = - \eta$ is sufficient. In terms of 16 dimensional gamma matrices and spinors this condition may be expressed as
\be
\g^{6789} \e \equiv \left( i \s^2 \otimes 1_{4} \otimes i \s^2 \right) \e = -\e,
\ee
with a similar expression for $\hat{\e}$. To impose this condition, we work with a basis of complex constant spinors of the form
\bea
\label{spinbasis}
\xi_a &=& \left( \begin{array}{c} 1 \\ i \end{array} \right) \otimes \chi_a \otimes   \left( \begin{array}{c} 1 \\ i \end{array} \right), \nn
\xi_{a+4} &=& \left( \begin{array}{c} 1 \\ -i \end{array} \right) \otimes \chi_a \otimes   \left( \begin{array}{c} 1 \\ -i \end{array} \right),
\eea
where $\chi_{a}$, $a=1,...4$ is a four-component spinor of the form $\chi_1 = (1,0,0,0)^t, \chi_2 = (0,1,0,0)^t$, etc. In total, this gives us a basis of eight spinors $\e_{a}$ which are adequate to describe our 16 component Weyl spinors subject to the  projection condition (\ref{adssproj}). To get the final Killing spinor one has to put a linear combination of these basis spinors in (\ref{killspin}) for the constant spinor $\e_0$.



In performing four fermionic T-dualities, a little trial and error with our basis of spinors reveals four commuting spinors
\bea
\label{spinors}
\mathcal{E}_1 &=& {\e}_1 + {\e}_6, \nn
\mathcal{E}_2 &=& {\e}_2 - {\e}_5, \nn
\mathcal{E}_3 &=& {\e}_3 + {\e}_8, \nn
\mathcal{E}_4 &=& {\e}_4 - {\e}_7,
\eea
which lead to an expression (\ref{C}) that may be integrated consistently. Here, ${\e}_i$ mean the Killing spinor one gets as a result of exponentiating the basis spinors in (\ref{spinbasis}). In other words,
\be
\label{expspin}
{\e}_i =  r^{-1/2}  e^{-\tfrac{\theta}{2} \g^{145} }  e^{\tfrac{\phi}{2} \g^{34}} e^{\tfrac{\psi}{2} \g^{45}} \xi_i.
\ee
Observe that the signs here are chosen to ensure that the fermionic isometries commute (\ref{commutespin}) and we have introduced $\CE_i$ to differentiate the commuting spinors from the basis spinors $\e_i$ . The resulting four-dimensional matrix $C_{ij}$ may then be expressed as
\be
\label{C1}
C = -\frac{16}{r}\left( \begin{array}{cccc}  i c_{\theta} + s_{\theta} s_{\phi} s_{\psi} & i s_{\theta} c_{\phi} & s_{\theta} s_{\phi} c_{\psi} & 0 \\ i s_{\theta} c_{\phi} & -i c_{\theta} + s_{\theta} s_{\phi} s_{\psi} & 0 & s_{\theta} s_{\phi} c_{\psi} \\ s_{\theta} s_{\phi} c_{\psi} & 0 & i c_{\theta} - s_{\theta} s_{\phi} s_{\psi}  & i s_{\theta} c_{\phi} \\  0 & s_{\theta} s_{\phi} c_{\psi} & i s_{\theta} c_{\phi} & - i c_{\theta} - s_{\theta} s_{\phi} s_{\psi} \end{array}\right),
\ee
where we have employed an obvious shorthand notation $s_{\theta} \equiv \sin \theta, c_{\theta} \equiv \cos \theta$, etc.

The shifted dilaton resulting from these four commuting fermionic T-dualities may then be calculated from (\ref{dilshift}). To do so, we can calculate the eigenvalues $\l$ of $C$
\be
\l = \frac{16}{r} \left(\pm s_{\theta} s_{\phi} \pm \sqrt{s_{\theta}^2 s_{\phi}^2-1}   \right),
\ee
where all combinations of the above signs give the four eigenvalues. Taking the logarithm of these eigenvalues and summing them one finds that
\be
\label{phishift}
\tilde{\phi} = \phi - 2 \ln r,
\ee
up to a constant. This is the first sign that everything is working out nicely. 

The change in the fluxes can then be read off from how the bispinor $F^{\a \b}$ transforms (\ref{fluxtrans}). In calculating $C_{ij}^{-1} \CE_i \otimes \hat{\CE}_j$ for the Killing spinors (\ref{spinors}), one finds a sparse matrix with constant entries that is antisymmetric. A cursory glance at (\ref{fluxbi}) reveals that one-form and five-form fluxes are symmetric while three-form fluxes are antisymmetric. Thus, the resulting fluxes using (\ref{spinors}) will all be three-forms. After a brute force calculation, one finds that
\be
\label{flux1}
C_{ij}^{-1} \CE_i \otimes \hat{\CE}_j = - \frac{i}{8} ( \g^{012} + \g^{345}) + \frac{1}{8} (\g^{279} - \g^{268}).
\ee
We immediately see that this expression is consistent with the projection conditions on $\CE_i$ under $\g^{6789}$, as the LHS side changes sign and so does the RHS when one makes use of $\g^{0136789} = - \g^{345}$. Inserting this expression back into (\ref{fluxtrans}) one finds the form of the fluxes post transformation. The resulting solution may now be written 
\bea
ds^2 &=& \frac{-dt^2 + dx^2 + dr^2}{r^2} + ds^2(S^3) + ds^2(T^4), \nn
F_3 &=& 2 i r dr \wedge J ,  
\eea
where ${J} \equiv dx^6 \wedge dx^8- d x^{7} \wedge dx^{9}$ is the K\"ahler form on $T^4$. Since $CY_2$ enjoys Hyper-K\"{a}hler structure, one has an $SU(2)$ of suitable K\"{a}hler forms, so we have the luxury of being a little loose. 
Observe that the original fluxes have cancelled leaving purely imaginary three-form fluxes. 

As a quick consistency check, we can confirm that the Bianchi, flux equations and Einstein equation are satisfied
\be
R_{\mu \nu} + 2 \nabla_{\mu} \nabla_{\nu} \phi = e^{2 \phi} \left( \frac{1}{2 (2!)} F_{\mu \s_1 \s_2} F_{\nu}^{~\s_1 \s_2} - \frac{1}{4 (3!)} g_{\mu \nu} F_{ \s_1 \s_2 \s_3} F^{\s_1 \s_2 \s_3} \right).
\ee

To establish self-duality of the original background, we now need to perform bosonic T-dualities along the $t$ and $x$ directions of $AdS_3$. As a result of the timelike T-duality the $i$ factor in the fluxes disappears \cite{Hull:1998vg} and they become real again while the dilaton shifts back to its original value. One now has a geometry describing D3-branes wrapped on a Kahler two-cycle in $CY_2$, however in this case the $CY_2$ is just $T^4$, so it corresponds to intersecting D3-branes. Finally, two T-dualities along say $x^6, x^8$ or $x^7,x^9$ in tandem with an inversion of the $AdS_3$ radial direction will then recover the original form of the geometry.  

In light of the observation that additional $T^4$ bosonic T-dualities are required to return the geometry to its original guise, it is worth taking time to wonder whether there is a combination of two $AdS_3$ bosonic T-dualities and four fermionic T-dualities that brings the geometry back on the nose. As the effect of two bosonic T-dualities is to convert a three-form flux into a one-form and five-form flux, immediately we recognise the need for the four fermionic T-dualities to source a one-form flux term. It is indeed easy to see that those terms cannot arise from the flux bispinor transformation. To see this expand $C_{ij}^{-1} \CE_i \otimes \hat{\CE}_j$ as 
\be
\label{fierz}
C_{ij}^{-1} \CE_i \otimes \CE_j \g^1 = \frac{1}{16} C_{ij}^{-1} \sum_{A} \CE_j \g^{1} \G_{A} \CE_i \G^{A},  
\ee
using a complete basis of sixteen dimensional gamma matrices $\G^{A} \in ~ \{ \mathbb{1}, \g^{\mu}, \dots \}$ and the Fierz identity. Now as $\CE_{i} \g^{1} \g^{\mu} \CE_j$ is antisymmetric when $\mu \neq 0, \mu \neq 1$, whereas $C_{ij}$ is symmetric, these terms automatically vanish meaning the the only one-forms that can appear are those proportional to $\g^0$ or $\g^{1}$. These latter possibilities are already precluded by the requirement that the spinors commute, so we see that one-form terms cannot be sourced during the transformation. This means that self-duality in $D=10$ and self-duality in the supercoset formulation in $D=6$ presented in \cite{Adam:2009kt,Dekel:2011qw} have a slightly different flavour.

\section{Complexified $S^3$}
Having shown that $AdS_3 \times S^3 \times T^4$ supported by RR three-form fluxes is self-dual, in this section we wish to consider T-dualities along the $S^3$. This approach is motivated by an alternative recipe for self-duality presented in \cite{fermTdual}. To this end we must complexify the geometry and perform bosonic T-duality w.r.t. these complex Killing directions. We will see that when we map these Killing directions back into our original coordinates (\ref{sphmet}) that the additional requirement that the Killing spinors commute with these isometries gives rise to an projection condition that selects the appropriate fermionic T-duality Killing spinors. 

Recall that we can think of the three-sphere as the surface in $\mathbb{R}^4$ given by 
\be
X_1^2+X_2^3+X_3^2+X_4^2=1.
\ee
For example if we choose the embedding coordinates to be 
\bea
 X_1=\cos \theta, &~& X_2=\sin \theta \cos\phi,\nn
X_3=\sin \theta \sin \phi \cos \psi,&~& X_4=\sin\theta \sin \phi\sin \psi,
\eea
we find the usual sphere metric (\ref{sphmet}). However if we complexify the embedding 
space to $\mathbb{C}^4$ then we can choose
\bea
 X_1=\frac{x_1}{w}, &~& X_2=\frac{x_2}{w},~\nn
X_3=\frac{i}{2w}(1-w^2 +x_1^2+x_2^2),&~& X_4=\frac{1}{2w}(1+w^2 -x_1^2-x_2^2),
\eea
where in particular $X^3$ is imaginary. Now the metric on the sphere becomes
\be
\label{spheremetcomp}
ds^2(S^3_{\mathbb{C}})=\frac{-dw^2+dx_1^2+dx_2^2}{w^2}.
\ee
This complexified $S^3$ we will refer to as de Sitter. Of course this transformation can be done without making reference to the embedding space
by making a complex coordinate transformation e.g.
\bea 
\label{comp_coord}
w&=& \frac{i}{\sin \theta \sin \phi} e^{-i \psi}, \nn
x_1 &=& \frac{i \cos \theta}{\sin \theta \sin \phi} e^{-i \psi}, \nn
x_2 &=& \frac{i \cos \phi}{\sin \phi} e^{-i \psi}. 
\eea


Using the change of coordinates (\ref{comp_coord}), one may rewrite the $\partial_{x_1}$ and $\partial_{x_2}$ Killing vectors of the analytically continued metric (\ref{spheremetcomp}) in terms of the original coordinates on the three-sphere as 
\bea
\partial_{x_1} &=& i e^{i \psi} \left( \sin \phi \partial_{\theta} + \cos \phi \cot \theta \partial_{\phi} + i \cot \theta \csc \phi \partial_{\psi} \right), \nn
\partial_{x_2} &=& i e^{i \psi} \left( \partial_{\phi} + i \cot \phi \partial_{\psi} \right).  
\eea

Allowing for two bosonic T-dualities on the $x_1$ and $x_2$ directions, we need to find fermionic directions commuting with these directions. Such a task boils down to finding Killing spinors whose Lie derivative with respect to $\partial_{x_i}~,i=1,2$ vanishes. Denoting the Killing vector $K$, the Lie derivative of a Killing spinor $\e$ with respect to the vector $K$ may be expressed as \cite{hep-th/9902066} 
\be
{\cal L}_{K} \eta = K^{M} \nabla_{M} \eta + \frac{1}{8} d K_{MN} \G^{MN} \eta.  
\ee 

Up to overall factors, a calculation reveals that both ${\cal L}_{\partial_{x_1}} \eta = 0 $ and ${\cal L}_{\partial_{x_2}} \eta = 0$ imply the single condition 
\be
\left[ i \g^{34} e^{\theta \g^{145} } - \cos \phi \g^{35}  e^{\theta \g^{145} } + \sin \phi \g^{45} \right] \e = 0,
\ee 
where in deriving this projection condition we have used $\hat{\e} = \g^1 \e$ and $ \g^{12345} \e = - \e$. In performing this calculation we have also used the Killing spinor equation (\ref{KSE}) and raised and lowered indices using the metric of the original three-sphere. Finally, inserting the explicit expression for $\e$ in terms of the constant spinor $\e_0$ from (\ref{killspin2}) and commuting the gamma matrices through one finds that $\e_0$ is subject to the projection condition 
\be
\g^{45} \e_0 = i \e_0. 
\ee

Returning to our original basis of eight Killing spinors (\ref{spinbasis}), one can form the following combinations obeying this additional projection condition: $\xi_1 - i \xi_3, \xi_2 - i \xi_4$, $\xi_5 - i \xi_7$ and $ \xi_6 - i \xi_8$. Exponentiating these constant spinors in a similar fashion to before as in (\ref{expspin}), one finds four commuting spinors, namely 
\bea
\CE_1 &=& \e_1 - i \e_3, \nn 
\CE_2 &=& \e_2 - i \e_4, \nn 
\CE_3 &=& \e_5 - i \e_7, \nn 
\CE_4 &=& \e_6 - i \e_8, 
\eea
and the resulting matrix 
\be
C = \frac{16}{r w}\left( \begin{array}{cccc}  0 &  0  & 0 & -1 \\ 0 & 0 & 1 & 0 \\ 0 & 1 & 0 & 0 \\ -1 & 0 & 0 & 0 \end{array}\right). 
\ee
Using this matrix $C$ and (\ref{dilshift}), the shift in the dilaton may be worked out 
\be
\label{dilshifts3}
\tilde{\phi} = \phi - 2 \ln ( r w). 
\ee

One can then proceed to work out the fluxes through the contracted term $C_{ij}^{-1} \CE_i \otimes \hat{\CE}_j$. As before, it is easy to check that this matrix is antisymmetric and therefore that the resulting fluxes will be three-forms. Also, one finds that the original fluxes are cancelled again. After some brute force computation, one may determine the form of the fluxes as a result of performing four commuting fermionic T-dualities. The resulting three-form flux may be written: 
\bea
F_3 =  - 2 i r dr \wedge dx_1 \wedge dx_2 + 2 i w dw \wedge dt \wedge d x,  
\eea
where we have written the result in terms of the de Sitter coordinates. 

Observe again that the fluxes are purely imaginary and will become real once we perform the timelike T-duality. The reader familiar with bosonic T-duality will also quickly take note that this flux term will revert to the original fluxes under T-duality along $t, x, x_1$ and $x_2$ directions. The dilaton shift (\ref{dilshifts3}) will be undone under this combination of bosonic T-dualities. Finally, an inversion in both $r$ and $w$ will bring us back to the original form of the geometry. 

In contrast to earlier work, we see here that one recovers the geometry without the need to perform T-dualities along $T^4$. 

\section{An adventure in S-duality}
Recall that type IIB supergravity \cite{hassan,Schwarz:1983qr} has an $SL(2,\mathbb{R})$ or S-duality action of the form
\be
\tau \rightarrow \frac{p \tau +q}{r \tau + s}, \quad \quad \left( \begin{array}{c} C^{(2)} \\ B \end{array} \right) \rightarrow \left( \begin{array}{cc} p & q \\ r & s \end{array} \right)\left( \begin{array}{c} C^{(2)} \\ B \end{array} \right),
\ee
where $\tau \equiv C^{(0)} + i e^{-\phi}$. Choosing $p = s = 0$, $ q = -r = 1$, we get an S-duality transformation where $C^{(2)}$ and $B$ are interchanged. We will now attempt to repeat the analysis of section 2. 

Before venturing further, we should state unequivocally that generically S-duality and fermionic T-duality do not commute. One can see this by starting with perhaps a D3-brane background with a five-form flux and performing fermionic T-dualities to turn on a $C^{(2)}$ potential which under S-duality will become a $B$-field. Reversing the process, no $B$-field is produced under $S$-duality and then the subsequent fermionic T-duality will not source one, so it is clear they do not commute. So from the offset, we may expect that a $B$-field could affect self-duality, but this is not obvious as under fermionic T-duality the $B$-field does not change and plays a spectator r\^{o}le. It is therefore instructive to work through an example.  

Firstly, it appears that the bosonic T-dualities along $AdS_3$ can no longer be done as they lead to singular metrics. As a quick illustration of the obstacle here, we attempt to do a timelike T-duality on $AdS_3$. From \cite{Hull:1998vg} we expect the RR fluxes to become pure imaginary while the $B$-field stays real and the usual Buscher rules apply. Inserting $B_{tx} = -\tfrac{1}{r^{2}}$ into the Buscher rule 
\be
\tilde{g}_{xx} = g_{xx} - \frac{1}{g_{tt}} \left( g_{tx}^2 - B_{tx}^2 \right), 
\ee 
one sees that $\tilde{g}_{xx}$ disappears after T-duality. In addition, the original $B$-field becomes a $\tilde{g}_{xt}$ cross-term in the metric. However, the damage is already done and the singularity in $g_{xx}$ prevents one from doing T-duality along $x$. TsT transformations \cite{Lunin:2005jy} also do not help to circumvent this obstacle. 

The difficulties above in performing bosonic T-duality along $AdS_3$ aside, one can still attempt to perform fermionic T-dualities. To that end, we determine the form the Killing spinors with $C^{(2)}$ replaced by $B$. The dilatino variation is the same and leads to the projector (\ref{adssproj}). The gravitino variation in our conventions is 
\be
\delta \Psi_{M} = \nabla_{M} \eta - \frac{1}{8} \G^{NP} H_{MNP} \s^3 \eta,  
\ee 
leading to the solution 
\be
\label{killspin3}
\eta = r^{-1/2} e^{\tfrac{\theta}{2} \G^{45} \s^3} e^{\tfrac{\phi}{2} \G^{34}} e^{\tfrac{\psi}{2} \G^{45}} \eta_0, 
\ee
where the constant spinor $\eta_0$ now satisfies a new Poincar\'e condition 
\be
\label{projB}
\G^{01} \s^3 \eta = - \eta ~~\Rightarrow \g^{1} \e = - \e, ~~\g^{1} \hat{\e} = \hat{\e}. 
\ee
Using these additional projection conditions, it is possible to show that imposing the commuting condition (\ref{commutespin}) also implies $\partial_{\mu} C_{ij} = 0$. In other words, we only have trivial constant transformation matrices $C_{ij}$ and not matrices that depend on the $AdS_3$ radial direction or $S^3$. To see this, we recall the Killing spinor projection conditions: 
\bea
&&\g^{1} \e = - \e, \quad \g^1 \hat{\e} = \hat{\e}, \nn
&& \g^{12345} \e = - \e, \quad \g^{12345} \hat{\e} = -\hat{\e}, \nn
&& \g^{6789} \e = -\e, \quad \g^{6789} \hat{\e} = - \hat{\e}, 
\eea
where the last two projection conditions are not independent. Using these projectors it is possible to show that the following spinor bilinears are zero:
\be
\e \g^{\mu} \e = \hat{\e} \g^{\mu} \hat{\e} = 0, ~~\mu = 2,\dots, 9. 
\ee
Furthermore, one can then show that the remaining commuting conditions, namely $\e \e + \hat{\e} \hat{\e} = \e \g^1 \e + \hat{\e} \g^{1} \hat{\e} = 0$ imply $\e \e = \hat{\e} \hat{\e} = 0$ using the projector in the first line.  In essence, the auxiliary matrix $C_{ij}$ describing the fermionic T-duality must be a constant and that the transformation is somewhat trivial in nature. 

As we have encountered a curious singularity in the $AdS_3$ bosonic T-dualities and observed that the fermionic T-duality is simply a constant rotation on the Killing spinors, it may be hoped that this singularity may be exorcised by considering the one-parameter family of S-dual solutions with zero dilaton. Returning to our S-duality formula and taking $p = s = \cos \a$, $r = - q = \sin \a$, one sees that $\tau$ remains unchanged under S-duality, so that the axion and dilaton remain zero (constant), while the effect of the T-duality is to turn on a $H = d B$. The presence of this B-field will then ensure that generically neither of the metric components $g_{tt}$ nor $g_{xx}$  disappear when one performs the bosonic T-dualities on $AdS_3$. Up to a rescaling of the radial direction $r$, one finds that the usual Poincar\'e writing of $AdS_3$ may be recovered after T-dualities in the $t$ and $x$ directions. This eliminates the obstacle of the singularity encountered previously and the only unknown concerns the fermionic T-duality. 

Recalling the Killing spinor equations, (\ref{dilatino}) and (\ref{KSE}), after some work one observes that the effect of the S-duality is simply to rotate the Killing spinors by a constant angle $\a$. Firstly, the dilatino variation leads to the projector 
\be
(\G^{012} + \G^{345} ) e^{-i \a \s^2} \eta =0, 
\ee
implying $\G^{012345} \eta = - \eta $ as before. The Poincar\'e projector on $AdS_3$ then becomes 
\be
\label{projCB}
\G^{01} \s^1 e^{i \a \s^2} \eta = \eta. \ee
Immediately, one appreciates that the previous projectors for a $C^{(2)}$ supported geometry (\ref{poinproj}) and a $B$-field supported geometry (\ref{projB}) are recovered when $\a = 0$ and $\a = \pi/2$ respectively. Finally, the gravitino variation along $S^3$ may be solved by 
\be
\tilde{\eta} = e^{i \tfrac{\alpha}{2} \s^2} \eta  =   r^{-1/2} e^{-\tfrac{\theta}{2} \G^{4 5} \s^1} e^{\tfrac{\phi}{2} \G^{3 4}} e^{\tfrac{\psi}{2} \G^{4 5}}\eta_0. 
\ee
As advertised, this is simply a rotation of the Killing spinor appearing in (\ref{killspin}). Complexifying the spinors appropriately, $\eta = \e + i \hat{\e}$, this is just the rotation on the Killing spinors under S-duality noted in \cite{Ortin:1994su}: 
\be 
\label{Sks}
\eta \rightarrow \exp \left( \frac{i}{2} \textrm{arg} (r \tau + s) \right) \eta. 
\ee

Employing this rewriting of the spinor in (\ref{projCB}), one finds the condition $\G^{01} \s^1 \tilde{\eta} = \tilde{\eta}$, meaning that the constant Majorana spinors appearing in $\eta_0$, i.e. $(\e_0, \hat{\e}_0)$, are simply related via $\hat{\e}_0 = \g^1 \e_0$. This allows us to then write 
\bea
\e &=& \left( \cos \tfrac{\a}{2} \cos \tfrac{\theta}{2} +  \sin \tfrac{\a}{2} \sin \tfrac{\theta}{2} \g^{45} -  \sin \tfrac{\a}{2} \cos \tfrac{\theta}{2} \g^1 - \cos \tfrac{\a}{2} \sin \tfrac{\theta}{2} \g^{145} \right) e^{\tfrac{\phi}{2} \g^{3 4}} e^{\tfrac{\psi}{2} \g^{4 5}} \e_0, \nonumber \\
\hat{\e} &=& \left( \cos \tfrac{\a}{2} \cos \tfrac{\theta}{2} \g^1 - \sin \tfrac{\a}{2} \sin \tfrac{\theta}{2} \g^{145} +  \sin \tfrac{\a}{2} \cos \tfrac{\theta}{2} - \cos \tfrac{\a}{2} \sin \tfrac{\theta}{2} \g^{45} \right) e^{\tfrac{\phi}{2} \g^{3 4}} e^{\tfrac{\psi}{2} \g^{4 5}} \e_0, \nonumber 
\eea
in terms of the single constant spinor $\e_0$. Then choosing the same spinors (\ref{spinors}), one finds four commuting fermionic isometry directions and that the matrix $C_{ij}$ appearing in (\ref{C1}) is unchanged up to an overall multiplying factor of $\cos \a$. In turn this means that the shifted dilaton becomes 
\be
e^{\tilde{\phi}} = \frac{\cos^2 \alpha}{r^2}. 
\ee
The fluxes resulting from these four fermionic T-dualities may be written 
\bea
\label{Sdualflux}
F_1 &=& 2 \frac{ \sin \a}{\cos^3 \a} r dr, \\ 
{F}'_3 &=& -2 \frac{ \sin^2 \a}{\cos^3 \a} r^2 \left[ \textrm{vol}(AdS_3) + \textrm{vol}(S^3) \right] +2  \frac{1}{\cos^3 \a} i r dr \wedge  {J} ,\\
F_5' &=& 2 \frac{\sin \a}{\cos^3 \a} (1+*) \biggl[  dt \wedge dx\wedge  \textrm{vol}(S^3) - i \frac{1}{r} dt \wedge dx \wedge dr  \wedge { J}\biggr], 
\eea
where ${J}$ is again the K\"{a}hler form on $T^4$ and we have used dashed notation to denote fluxes incorporating $H$. Observe that when $\a = 0$, we recover our results from the previous section with both $F_1$ and $F_5$ vanishing, whereas when $\a  =\pi/2$, the dilaton shift develops a singularity. 

In addition, this is the only example in this paper where the matrix $C_{ij}^{-1} \CE_i \otimes \hat{\CE}_j$ is a blend of both symmetric and antisymmetric matrices. In the other examples involving $AdS_3 \times S^3 \times T^4$, this matrix is antisymmetric while in the seminal $AdS_5 \times S^5$ example presented in \cite{fermTdual} this matrix is symmetric. 

We also take note of the fact that both real and imaginary fluxes appear. Since timelike T-duality exchanges real and imaginary fluxes, in order to extract a real solution, we require some cancellations to happen through the process of taking T-duality along both the $t$ and $x$-directions. We begin by performing T-duality on the NS sector of the solution by first performing a T-duality along the $x$-direction and then the $t$-direction. The usual Buscher rules lead to the geometry 
\bea
ds^2 &=& \frac{r^2}{\cos^2 \a} (-dt^2 + dx^2) + \frac{dr^2}{r^2} + ds^2(S^3) + ds^2(T^4), \nn
\tilde{\phi} &=& \phi + \frac{1}{2} \ln \left( \frac{r^2}{\cos \a} \right), \nn
H &=& - 2 \frac{\sin \a}{\cos^2 \a} r dt \wedge dx \wedge dr + 2 \sin \a \textrm{vol}(S^3).  
\eea
Up to a rescaling in $r$, we see that $AdS_3$ in Poincar\'e may be recovered. However this rescaling does not go through to the dilaton shift. This is because there is a missing factor of $\cos \a$ since fermionic T-duality gives rise to two, while bosonic T-duality only generates a $\cos \alpha$ factor after the initial T-duality. 

Applying the T-duality rules to the RR fluxes one finds:
\bea
F_3 &=& 2 i \frac{\sin \a}{\cos^3 \a} \left[ - r dt \wedge dx \wedge dr + \textrm{vol}(S^3) \right], \nn
F_5 &=& 2 \frac{1}{\cos^3 \a} (1+ *) \left[ r dt \wedge dx \wedge dr \wedge {J}\right]. 
\eea

For completeness we have exhibited the final solution. While this calculation may be regarded as a success in the sense that the fermionic T-duality works out nicely, we see that a real background can only be recovered in the limit $\a \rightarrow 0$, in which case the $B$-field disappears. Thus the $B$-field acts as an obstacle to self-duality and it is expected that the fundamental reason is that S-duality and fermionic T-duality fail to commute. Just as bosonic T-duality and S-duality have to be enlarged to U-duality \cite{Hull:1994ys}, it is certainly possible to combine fermionic T-duality and S-duality into  a larger fermionic ``U-duality". At the level of the Killing spinors this can work by combining the transformation of (\ref{Sks}) with the rotation of the Killing spinors coming from fermionic T-duality, with the obvious caveat being that one would generate a complex supergravity solution. Finding real solutions requires more work which we postpone for future work. 

\section{Comments on $K3$} 
\label{sec:K3}
Earlier work strongly suggests that fermonic T-duality in the background $AdS_3 \times S^3 \times CY_2$ cares little about the internal space. Here we comment on why this is the case. Recall from (\ref{C}) that the right hand side is a vector bilinear. Using the projection condition $\g^{6789} \e = -\e$, it is easy to show that all vector bilinears on $CY_2$ vanish. As such, the matrix $C_{ij}$ has to be independent of $CY_2$. 

Now although $C_{ij}$ does not depend on $CY_2$, a cursory glance at (\ref{fluxtrans}) will show that the transformation appears to depend on the explicit form of the Killing spinors. These are only covariantly constant along $K3$ and not constant as in the case of $T^4$, so naively they have some $K3$ dependence.  One can again turn to Fierz identities and expand in terms of one-forms, three-forms and five-forms as was suggested in (\ref{fierz}). We begin by looking at the one-forms. 

Simply from symmetries one can infer that all one-forms vanish. Though we do not know the spinors, if we assume they commute and satisfy (\ref{commutespin}) then we see that neither $\g^0$ nor $\g^1$ can appear as one-forms. On top of that, none of $\g^{i}$, $i=2,\dots,9$ can appear as $\g^{1} \g^{i}$ is antisymmetric and $C_{ij}$ is symmetric in the indices. This kills off the prospect of one-forms. We focus on the three and five-forms. 

Returning to (\ref{fierz}), it is a bispinor that satisfies the same projection conditions as the constituent spinors. This means that when it is expanded via Fierz that the gamma matrices have to appear so that acting with $\g^{012345}$ leads to a minus sign. Some trial and error reveals that $\g^{01i}$, $i=2,...,5$ must appear together, otherwise the bilinears constructed will not be symmetric as one will not get a one-form and a four-form\footnote{With our choice of gamma matrices $\g^0$ is simply the identity.}. Now, showing that only $\g^{012}$ can appear involves using the dilaton shift (\ref{phishift}) which is derived from $C_{ij}$ and depends only on $r$, which in our vielbein (\ref{vielbein}),  corresponds to $e^2$. Now the $\g^{012}$ term in the Fierzed expansion has a coefficient 
\be
-\frac{1}{16} \CE_i \g_2 \CE_j C_{ij}^{-1} \g^{012} =  \frac{i}{32} \partial_{2} C_{ij} C_{ij}^{-1} = \frac{i}{32} \partial_2 \textrm{Tr} \log C, 
\ee
where we have used (\ref{C}). Now we can simply use (\ref{phishift}) to get the term $-\frac{i}{8} \g^{012}$ which indeed is the expression appearing in (\ref{flux1}). Acting with $\g^{12345}$ we can then generate the coefficient in from the $\g^{345}$ in the same expression. 

Having gone through some details, it should be clear that a  $\g^{013}$ term cannot appear as $ \textrm{Tr} \log C$ does not depend on the $\theta$ coordinate. Similar reasoning holds for the other expressions. The terms involving the K\"ahler form from $CY_2$ can be determined in a similar fashion while similar arguments can be used to figure out that five-forms do not appear. 

So, what have we learned? We have seen that the internal $CY_2$ completely drops out. In short,  $C_{ij}$ cannot depend on $CY_2$ and since $C_{ij}$ generates the shift in the dilaton and the shift in the fluxes, once one rewrites the latter, one discovers that fermionic T-duality does not care too much about the particular $CY_2$. So, we are led to the conclusion that the D1-D5 near-horizon is self-dual in a pretty general way.  

\section*{Acknowledgements} 
We have enjoyed discussing fermionic T-duality with a host of interesting people: Ido Adam, Ilya Bakhmatov, David Berman, Marco Bianchi, Volker Braun, Chang-Young Ee, Jerome Gauntlett, Jan Gutowski, Bernard Julia, Yolanda Lozano, Patrick Meessen, Werner Nahm, Hiroaki Nakajima, Yaron Oz, Soojong Rey, Hyeonjoon Shin \& Dan Waldram. In particular we would like to thank Tristan McLoughlin for frequent input, sharing his curiosity about $AdS_4 \times \mathbb{C} \textrm{P}^3$ self-duality and his continued enthusiasm. We would like to acknowledge the kind hospitality of the Dublin Institute for Advanced Study and the Isaac Newton Institute, Cambridge where the \textit{Mathematics and Applications of Branes in String and M-Theory} program has seeded many stimulating discussions. 
 
\appendix
\section{Conventions}
Throughout this work we employ the real representation for the ten-dimensional gamma matrices appearing in the Clifford algebra Cl$(9,1)$. We choose our gamma matrices to be
\be
\label{gammadec}
\G^0 = i\sigma_2 \otimes \mathbb{1}_{16}, \quad \G^i = \sigma_1 \otimes \g^i,
\ee
where we further decompose
\be
\begin{array}{ccccccc}
\g^1 = \sigma_2 & \otimes & \sigma_2 & \otimes & \sigma_2 & \otimes & \sigma_2, \\
\g^2 = \sigma_2 & \otimes & 1             & \otimes & \sigma_1 & \otimes & \sigma_2, \\
\g^3 = \sigma_2 & \otimes & 1             & \otimes & \sigma_3 & \otimes & \sigma_2, \\
\g^4 = \sigma_2 & \otimes & \sigma_1 & \otimes & \sigma_2 & \otimes & 1,\\
\g^5 = \sigma_2 & \otimes & \sigma_3 & \otimes & \sigma_2 & \otimes & 1,\\
\g^6 = \sigma_2 & \otimes & \sigma_2 & \otimes & 1             & \otimes & \sigma_1, \\
\g^7 = \sigma_2 & \otimes & \sigma_2 & \otimes & 1             & \otimes & \sigma_3, \\
\g^8 = \sigma_1 & \otimes & 1             & \otimes & 1              & \otimes & 1, \\
\g^9 = \sigma_3 & \otimes &1  &\otimes & 1 & \otimes & 1.
\end{array}
\ee
Observe here that $\g^9 = \g^1 \cdots \g^8$. Our gamma matrices may be written in terms of 16 dimensional blocks as
\be
\label{16dblock}
\G^{\mu} = \left( \begin{array}{cc} 0 & (\g^{\mu})^{\a \b} \\ \g^{\mu}_{\a \b} & 0 \end{array}\right), ~~C = \left( \begin{array}{cc} 0 & c_{\a}^{~\b} \\  \bar{c}^{\a}_{~\b} & 0 \end{array}\right),~~\G^{11} = \left( \begin{array}{cc} \delta^{\a}_{~ \b} & 0 \\ 0 & \delta_{\a}^{~\b} \end{array}\right),
\ee
where the indices take values $\a, \b = 1,\cdots 16$.

Under the ten-dimensional chirality operator $ \G^{11} = \G^{0} \cdots \G^9  = \s_3 \otimes \mathbb{1}_{16}$, we have two inequivalent 16 component Weyl spinors $\psi_{\pm}$ satisfying $\G^{11} \psi_{\pm} = \pm \psi_{\pm}$. Working in a Majorana representation where $C = \G^0$, and
\bea
C \G^{\m} C^{-1} = - \G^{\m t}, \nn
C^t = - C,
\eea
we see that further imposing the Majorana condition on $\psi_{\pm}$ results in them being real.

\section{$(6,4)$-split and Killing spinors} 
Here, by way of a supplement, we derive the Killing spinors making use of a reduction on the internal $CY_2$. We consider the following split of the $D=10$ gamma matrices 
\bea
\label{eq:64decomp}
\G_{\mu} &=& \rho_{\mu} \otimes \g_4, \nn
\G_m &=& 1 \otimes \g_m, 
\eea
where $\rho_{\mu}, ~\mu=0,\dots,5$ and $\g_m, ~m=6,\dots,8$, denote $D=6$ and $D=4$ gamma matrices respectively and we also define $\g_4 \equiv \g_{6789}$.  

We begin by adopting a consistent set of conventions \cite{Sohnius:1985qm}, where we have in dimension $D$, the following intertwiners
\bea
A \G_{M} A^{-1} &=& \G_M^{\dagger}, \nn
C^{-1} \G_{M} C &=& - \G_{M}^{T}, \nn
D^{-1} \G_M D &=& -\G_M^{*},   
\eea 
where for consistency $D = C A^{T}$. We take all $A$ to be Hermitian and $C_{6}$ is symmetric whereas $C_{10}$ and $C_4$ are antisymmetric. It is also possible to define another intertwiner $\tilde{D} = \G_{D+1}^{-1} D$. Defining a conjugate spinor $\e^c$ by 
\be
\e^c = D \e^*, 
\ee
we see that the existence of Majorana spinors, i.e. those satisfying $\e^c = \e$, depends on the properties of $D$ or $\tilde{D}$. Irrespective of whether one uses $D$ or $\tilde{D}$, we have the following \cite{Sohnius:1985qm}
\bea
\label{Dint}
D_{10}^{*} &=& D_{10}^{-1}, \nn
D_6^* &=& - D_{6}^{-1}, \nn
D_{4}^{*} &=& -D_4^{-1}, 
\eea
meaning that Majorana spinors can only be defined in $D=10$. 
The intertwiners may be decomposed as  
\bea
A_{10} &=& A_6 \otimes A_4, \nn
C_{10} &=& C_6 \otimes C_4, \nn
D_{10} &=& D_6 \otimes D_4. 
\eea
For simplicity, we take $A_4 = 1$, so that $\g_m^{\dagger} = \g_m$ and $\g_4^{\dagger} = \g_4$. We then have from (\ref{Dint}) that 
\bea
C_6^* A_6^{\dagger} &=& - A_6^{-T} C_6^{-1}, \nn
C_{4}^{*} &=& - C_4^{-1}. 
\eea 

As IIB supergravity is parameterised by two Majorana-Weyl Killing spinors, $\e_i$, $i=1,2$, it suffices to write them as 
\be
\e_i = \chi_i \otimes \eta + \chi_i^{c} \otimes \eta^c, 
\ee
where $\chi_i^c = D_6 \chi_i^*$ and $\eta^c = C_4 \eta^*$ and $(\chi_i^c)^c = - \chi_i, ~(\eta^c)^c = - \eta$ follows from (\ref{Dint}). Observe that $\e_i$ are Majorana through construction and the $D=10$ Weyl condition is satisfied provided 
\be
\label{projk3}
\rho^{012345} \chi_i = - \chi_i, \quad \g_4 \eta = - \eta, 
\ee
where the dilatino variation plays a hand in picking out the minus sign in each case. The properties of $D_6$ and $C_4$ mean that the conjugate spinors $\chi_i^c$ and $\eta^c$ satisfy identical projection conditions as their respective spinors. Finally, the $D=10$ Killing spinor can be obtained by complexifying $\e = \e_1 + i \e_2$, resulting in 
\be
\e = \xi_1 \otimes \eta + \xi_2^c \otimes \eta^c. 
\ee
where $\xi_1 = \chi_1 + i \chi_2$ and $\xi_2 = \chi_1 - i \chi_2$. The conjugate spinor which will appear later in the Killing spinor equations is then 
\be
\e^c =  \xi_2 \otimes \eta + \xi_1^c \otimes \eta^c.
\ee

Having already imposed the dilatino variation through adopting the minus sign in (\ref{projk3}), it remains to satisfy the gravitino variation. In terms of $\e$ and $\e^c$, it may be rewritten as  
\be
\delta \Psi_{M} = \nabla_M \e - \frac{i}{3! 8} e^{\phi} \G^{KLN} F_{KLN} \G_M \e^c = 0.  
\ee
Inserting in the earlier expressions for $\e$ and $\e^c$, one finds that the Poincar\'e Killing spinors on $AdS_3$ can be extracted: 
\bea
\xi_i &=& r^{-1/2} \tilde{\xi}_i, \nn \tilde{\xi}_2 &=& i \g^{01} \tilde{\xi}_1. 
\eea
The latter relationship implies $\chi_2 = - \rho^{01} \chi_1$, so the two Majorana-Weyl spinors are related. The Killing spinor equation on $S^3$ can be most easily solved by defining 
\be
\zeta = \left( \begin{array}{c} \xi_1 \\ \xi_2 \end{array} \right). 
\ee
The Killing spinor equation may then be solved leading to 
\be
\zeta = r^{-1/2} e^{-\frac{\theta}{2} \rho^{45} \s^2} e^{\frac{\phi}{2} \rho^{34}} e^{\frac{\psi}{2} \rho^{45}} \zeta_0,
\ee
where $\zeta_0$ is a constant spinor and the Pauli matrix $\s^2$ acts in a natural way. Note, we also have the KSE along $CY_2$, however this is simply solved through demanding $\eta$ is covariantly constant, $\nabla_{m} \eta =0$.

\end{document}